\documentclass[preprint,12pt]{elsarticle}

\usepackage{amssymb}
\usepackage{color}

\journal{Nuclear Physics A}

\def\bea {\begin{eqnarray}}
\def\eea {\end{eqnarray}}

\def\be {\begin{equation}}
\def\ee {\end{equation}}
\def\nn {\nonumber}

\newcommand{\ep}{\epsilon}

\newcommand{\om}{\omega}

\newcommand{\F}{F_\pi}

\newcommand{\J}{J/\psi}

\begin{document}

\begin{frontmatter}

\title{Diffusion of hidden charm mesons in hadronic medium}
\author[label1]{Sukanya Mitra}
\author[label2]{Sabyasachi Ghosh}
\cortext[mycorrespondingauthor]{Corresponding author}
\ead{sabyaphy@gmail.com}
\author[mymainaddress,mysecondaryaddress]{Santosh K. Das}
\author[label1]{Sourav Sarkar}
\author[label1]{Jan-e Alam}
\address[label1]{Theoretical Physics Division, 
Variable Energy Cyclotron Centre, 1/AF, Bidhan Nagar, 
Kolkata - 700064}
\address[label2]{Instituto de Fisica Teorica, Universidade Estadual Paulista, 
Rua Dr. Bento Teobaldo Ferraz, 271, 01140-070 Sao Paulo, SP, Brazil}
\address[mymainaddress]{Department of Physics and Astronomy, University of Catania, 
Via Santa Sofia 64, 1-95125 Catania, Italy}
\address[mysecondaryaddress]{Laboratori Nazionali del Sud, INFN-LNS, 
Via Santa Sofia 62, I-95123 Catania, Italy}

\begin{abstract}
The drag and diffusion coefficients of a hot hadronic medium 
have been evaluated by using hidden charm 
mesons as probes. The scattering amplitudes required  for the evaluation of these coefficients 
are calculated using an effective theory and   scattering
lengths obtained from lattice QCD calculations. 
It is found that although the magnitude of the transport coefficients are small their temperature variation is 
strong.  The insignificant momentum diffusion of $J/\psi$ in the hadronic medium keeps their momentum
distribution largely unaltered.  Therefore, the task of characterization of quark gluon plasma 
by using the observed suppression of $J/\psi$ at high momentum will be comparatively easier.
\end{abstract}

\begin{keyword}
Drag and Diffusion coefficients, Hidden charm meson, Heavy Ion Collision.
\end{keyword}

\end{frontmatter}


\section{Introduction}
The experimental evidence of $J/\psi$ suppression
by NA50~\cite{NA50}, NA60~\cite{NA60} as well as by the PHENIX~\cite{Phenix} 
collaboration has long been suggested as a signal of quark-gluon plasma (QGP)
formation in heavy ion collisions~\cite{Satz}. However,
other mechanisms such as the $J/\psi$ absorption by comoving hadrons 
have also been proposed as an alternative mechanism to explain 
the suppression~\cite{Vogt}, indicating that 
the inelastic scattering rates of $J/\psi$
in the hadronic phase is significant 
~\cite{Haglin,Liu_Ko, Oset}. In addition, the 
opening of $J/\psi\rightarrow D{\overline {D}}$ decay in the 
medium due to in-medium modification of D mesons~\cite{Friman_D,Ghosh_D_NPA}
may also play a significant role in $J/\psi$ suppression 
in a hadronic environment. 

Heavy quark transport in hadronic matter is a topic of high contemporary interest 
~\cite{Laine,Rapp,SSSJ_D,Juan_D,SSSJ_B,Juan_B, Juan_RAA}. The drag and
diffusion of open charm~\cite{SSSJ_D} and bottom~\cite{SSSJ_B} mesons and the
role of hadronic matter in their suppression in heavy ion
collisions~\cite{SSSJ_RAA} have been investigated using effective hadronic
interactions based on heavy quark effective theory. The
suppression of heavy flavors in the hadronic phase in comparison to QGP was found to be 
smaller at LHC than at RHIC, suggesting that the characterization of QGP
at LHC would be less complicated than at RHIC ~\cite{SSSJ_RAA}.  

Recently we have obtained the
drag and diffusion of the $\Lambda_c$ baryon in hadronic matter~\cite{Lamdac}  
(See also Ref.~\cite{Juan_L}) and found those to be significant.  In fact, the
drag of the $\Lambda_c$ being lower than that of the $D$ mesons was seen to 
non-trivially affect the $p_T$ dependence  of the $\Lambda_c/D$ ratio and thus the $R_{AA}$ 
of single electrons originating from the decay of $\Lambda_c$. Motivated by these
results we proceed to study the 
temperature variation of the drag and diffusion coefficients of 
$J/\psi$ and $\eta_c$ in a comoving hadronic medium. 
For evaluating these  quantities the required 
interaction cross sections have been evaluated employing an
effective hadronic Lagrangian. Drag and diffusion coefficients have also been
estimated using T-matrix elements extracted from scattering lengths obtained
from lattice QCD calculations.

In the next section we provide the formulae for the drag and diffusion
coefficients followed by a discussion on the matrix elements
of elastic scattering of the $J/\psi$ with the light vector mesons in section III.
Results are given in Section IV and finally a summary in Section V. We provide
the squared matrix elements in the appendix.  

\section{Formalism}
The drag ($\gamma$) 
and diffusion ($D$) coefficients of $J/\psi$ and $\eta_c$ are obtained
from  the elastic scattering of $J/\psi$
with the light thermal hadrons ($H$) which constitute the equilibrated thermal 
medium. 
For the process $J/\psi,\eta_c(p_1) + H(p_2) \rightarrow J/\psi,\eta_c(p_3) + H(p_4)$,
the drag $\gamma$ can be expressed as~\cite{BS}:
\begin{equation}
\gamma=p_iA_i/p^2~,
\end{equation}
where $A_i$ is given by 
\bea
A_i&=&\frac{1}{2E_{p_1}} \int \frac{d^3p_2}{(2\pi)^3E_{p_2}} \int \frac{d^3p_3}{(2\pi)^3E_{p_3}}
\int \frac{d^3p_4}{(2\pi)^3E_{p_4}}  
\frac{1}{g_{(J/\psi,\eta_c)}} 
\nonumber \\
&&\sum  \overline{|M|^2} (2\pi)^4 \delta^4(p_1+p_2-p_3-p_4) 
{f}(p_2)\{1\pm f(p_4)\}[(p_1-p_3)_i] 
\nn\\
&\equiv& \langle \langle
(p_1-p_3)\rangle \rangle~.
\label{eq1}
\eea
The $g_{(J/\psi,\eta_c)}$ is the statistical degeneracy of the probes, 
$J/\psi$ or $\eta_c$.
The thermal distribution function $f(p_2)$ of the hadron $H$ in the incident
channel takes the form of Bose-Einstein or Fermi-Dirac distribution
depending on its spin and 
$1\pm f(p_4)$ are their corresponding Bose enhanced or Pauli blocked
phase space factor in their final states.
The drag coefficient of Eq.(\ref{eq1}) is just a
measure of the thermal average of the  
momentum transfer, $p_1-p_3$
weighted by the square of the invariant amplitude
$\overline{\mid M\mid^2}$ for the elastic scattering of $J/\psi$ and $\eta_c$
with thermal hadrons, generically denoted as $H$. 

In a similar way, the diffusion coefficient $D$ can be defined as:
\begin{equation}
D=\frac{1}{4}\left[\langle \langle p_3^2 \rangle \rangle -
\frac{\langle \langle (p_1\cdot p_3)^2 \rangle \rangle }{p_1^2}\right]~.
\label{diffusion}
\end{equation}

With an appropriate choice of $T(p_3)$
both the $\gamma$ and $D$ can be obtained from
a single expression, which is given by
\bea
\ll T(p_1)\gg&=&\frac{1}{512\pi^4} \frac{1}{E_{p_1}} \int_{0}^{\infty} 
\frac{p_2^2 dp_2 d(cos\chi)}{E_{p_2}} 
\hat{f}(p_2)\{ 1\pm f(p_4)\}\frac{\lambda^{\frac{1}{2}}(s,m_{p_1}^2,m_{p_2}^2)}{\sqrt{s}} 
\nonumber \\ 
&&\int_{1}^{-1} d(cos\theta_{c.m.})
\frac{1}{g} \sum  \overline{|M|^2} \int_{0}^{2\pi} d\phi_{c.m.} T(p_3)~,
\label{transport}
\eea
where $\lambda(s,m_{p_1}^2,m_{p_2}^2)=(s-m_{p_1}^2-m_{p_2}^2)^2-4m_{p_1}^2m_{p_2}^2$,
is the triangular function.
\section{Dynamics}

In this section we will discuss the two body elastic scattering of $J/\psi$ with
the constituent hadrons in the medium within the framework of SU(4) symmetry.
Some of the relevant features of these interactions are mentioned below and for details 
we refer to~\cite{Haglin, Liu_Ko}
to avoid repetition. The aim of the present work  is to estimate  the momentum diffusion 
coefficient of hot hadrons produced as a result 
of  phase transitions from an expanding QGP formed in  heavy ion collisions at relativistic energies.
That is we would like to understand how efficiently the momentum of the $J/\psi$ propagating 
through a hadronic medium is transferred to the medium  enabling to estimate momentum
diffusion or shear viscous coefficient of the medium~\cite{sm}.  Since $J/\psi$ is used here as a probe
to extract the coefficient of momentum diffusion for characterizing the medium,
its detection in the final reaction channel is essential. 
Therefore, the inelastic processes which kill the $J/\psi$  are not considered here. 
For this reason, we consider only the  two body elastic processes {\it e.g.} 
$h_1 + J/\psi\rightarrow h_1 +  J/\psi$, in the present work ($h_i$'s are hadrons in the medium).
Although, $J/\psi$ appears in the final channel in the processes like  $h_1 + J/\psi\rightarrow h_2 + h_3+J/\psi$, 
analogous to the gluon
bremsstrahlung by heavy quarks in the QGP phase,  are also ignored, 
as their contributions  may be smaller than the two body elastic processes.

The hot hadronic matter produced in the later stages of relativistic heavy ion
collisions is populated by light pseudo-scalars and vector mesons like $\pi, K, \eta, \rho^0, \omega$ and
$\phi$. The magnitude of hadronic scatterings are estimated 
either by introducing different perturbative or non-perturbative approach at 
quark level~\cite{Peskin,Martins} or by using an effective Lagrangian 
to calculate Feynman diagrams. Concerning the latter approach, the SU(4) is
the smallest possible symmetry group which includes the charmonium state 
explicitly along with the light and heavy pseudo-scalar and vector mesons.
The corresponding pseudo-scalar and vector meson matrices as
well as the chiral Lagrangian are given in Refs.~\cite{Haglin,
Liu_Ko} which can be readily used for the present purpose 
of evaluating the drag and diffusion coefficients of
$J/\psi$ in hadronic matter. 
Since SU(4) symmetry is badly broken by the large mass of the
charmed meson, terms involving hadron masses are included in the chiral 
Lagrangian using the experimentally determined values of SU(4) model parameters. 

Since pions
are identified with the  Nambu-Goldstone bosons of QCD
their interaction strength with other hadrons should abruptly decrease in the chiral limit. 
We recall the standard relation~\cite{Weinberg,Hatsuda} 
for the s-wave scattering length of pion with a heavy meson like  $J/\psi$,
\begin{equation}
a^{\pi J/\psi}_{l=0}=-(1+\frac{m_\pi}{m_{J/\psi}})^{-1}
\frac{m_\pi}{4\pi\F^2}{\vec I}_\pi\cdot{\vec I}_{J/\psi} +{\cal O}(m_{\pi}^2),
\label{scat_pion}
\end{equation}
with the dot product defined as, 
\begin{equation}
{\vec I}_\pi\cdot{\vec I}_{J/\psi}=\frac{1}{2}[I(I+1)-I_{J/\psi}(I_{J/\psi}+1)-I_\pi(I_\pi+1)].
\end{equation}
$I_\pi, I_{J/\psi}$ are respectively the isospin quantum numbers of the pion and $J/\psi$ and $I$ represents
their total isospin quantum number. In the chiral limit ($m_\pi\rightarrow 0$), 
the first term of eq.(\ref{scat_pion}) exactly vanishes. When the other hadron is $J/\psi$ or $\eta_c$, 
this term in (\ref{scat_pion}) vanishes exactly, not only in the chiral limit but also 
for finite pion mass (because ${\vec I}_\pi\cdot{\vec I}_{J/\psi}=0$). Hence the 
contribution of the pion in the charmonium scattering length starts from ${\cal O}(m_{\pi}^2)$, 
which indicates that at least at low energy the pion-charmonium interaction  is
weak. This is in confirmation with the results obtained using the meson exchange model
of Haglin et.al ~\cite{Haglin}, where it was found that the elastic 
channels of $J/\psi$ interaction involving the light pseudo-scalars are significantly smaller
in comparison with the vector mesons. They have found that  $\pi$, $\eta$, and $K$
elastic cross sections with $J/\psi$ are of order 100 fb, 1 nb and 100 nb respectively.
On the other hand the contributions for elastic scattering with $\rho^0,\omega$ and
$\phi$ mesons are quantitatively much larger, up to about a few mb. 
Hence, the elastic scattering  of the
heavy charmonium states like $J/\psi$ and $\eta_c$ with 
vector mesons are considered here. These processes involve vector-vector-pseudo-scalar interactions which are not 
present in the chiral Lagrangian.
The relevant effective interaction describing 
$J/\psi +V\rightarrow \eta_c\rightarrow J/\psi +V$ processes~\cite{Haglin} is
\be
{\cal L}_{JV\eta_c}=g_{JV\eta_c}\ep_{\alpha\beta\sigma\delta}\{\partial^{\alpha}J/\psi^{\beta}\}
\{\partial^{\sigma}V^{\delta}\}\eta_c~,
\label{Lag}
\ee
where $g_{JV\eta_c}=2.44$ GeV$^{-1}$, $7.03$ GeV$^{-1}$ and $4.51$ GeV$^{-1}$
for $V=\rho^0, \om$ and $\phi$ respectively \cite{Haglin}. 

\begin{figure}
\begin{center}
\includegraphics[scale=0.5]{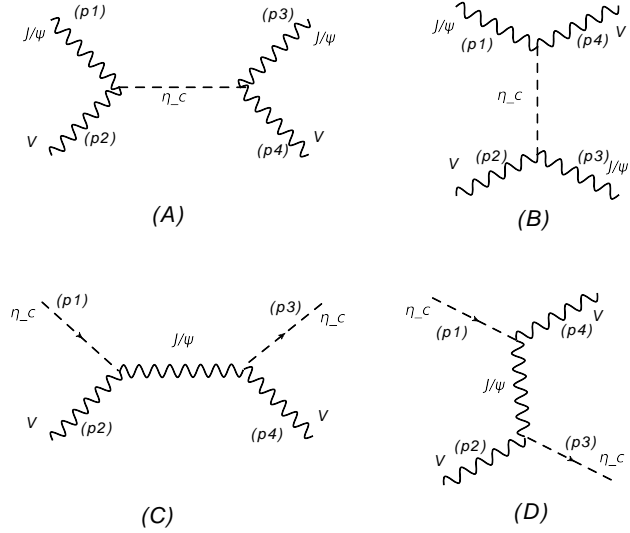}
\caption{The s and u channel of $J/\psi$-$V$ scattering
via $\eta_c$ exchange are respectively depicted in
diagrams (A) and (B). Diagrams (C) and (D) are the same
for the $\eta_c$-$V$ scattering via intermediary $J/\psi$.}
\label{jpsi_scat}
\end{center}
\end{figure}

The $s$ and $u$ channel diagrams for the process
$J/\psi +V\rightarrow \eta_c\rightarrow J/\psi +V$ 
are shown in the panels (A) and (B) of Fig.\ref{jpsi_scat}.
The matrix elements for the two channels are respectively
given by,
\be
M_s=-g_{JV\eta_c}^2[ \varepsilon^\beta(p_1)\varepsilon^\delta(p_2)
\varepsilon^{*\beta_1}(p_3)\varepsilon^{*\delta_1}(p_4)
\epsilon_{\alpha\beta\sigma\delta}~p_1^\alpha p_2^\sigma~
\epsilon_{\alpha_1\beta_1\sigma_1\delta_1}~p_3^{\alpha_1} p_4^{\sigma_1}]
/(s-m_{\eta_c}^2)
\ee
and
\be
M_u=-g^2_{JV\eta_c}[ \varepsilon^\beta(p_1)\varepsilon^{*\delta}(p_4)
\varepsilon^{\delta_1}(p_2)\varepsilon^{*\beta_1}(p_3)
\epsilon_{\alpha\beta\sigma\delta}~p_1^\alpha p_4^\sigma~
\epsilon_{\alpha_1\beta_1\sigma_1\delta_1}~p_2^{\sigma_1} p_3^{\alpha_1}]
/(u-m_{\eta_c}^2).
\ee

The $s$ and $u$ channel diagrams of the $\eta_c$ meson
scattering with the thermalized vector mesons by exchanging $J/\psi$
are shown in the panels (C) and (D) of Fig.~\ref{jpsi_scat}.
The respective matrix elements  are given by
\bea
M_s&=&-g^2_{JV\eta_c}[ \varepsilon^\delta(p_2)\varepsilon^{*\delta_1}(p_4)
\epsilon_{\alpha\beta\sigma\delta}(p_1+p_2)^\alpha p_2^\sigma
\epsilon_{\alpha_1\beta_1\sigma_1\delta_1}(p_1+p_2)^{\alpha_1} p_4^{\sigma_1}]
\nn\\
&&\{-g^{\beta\beta_1}+\frac{(p_1+p_2)^\beta(p_1+p_2)^{\beta_1}}{m_J^2}\}/
(s-m_J^2)
\eea
and
\bea
M_u&=&-g^2_{JV\eta_c}[ \varepsilon^{*\delta}(p_4)\varepsilon^{\delta_1}(p_2)
\epsilon_{\alpha\beta\sigma\delta}(p_1-p_4)^\alpha p_4^\sigma
\epsilon_{\alpha_1\beta_1\sigma_1\delta_1}(p_1-p_4)^{\alpha_1} p_2^{\sigma_1}]
\nn\\
&&\{-g^{\beta\beta_1}+\frac{(p_1-p_4)^\beta(p_1-p_4)^{\beta_1}}{m_J^2}\}/
(u-m_J^2)~.
\eea

\begin{figure}
\begin{center}
\includegraphics[scale=0.3]{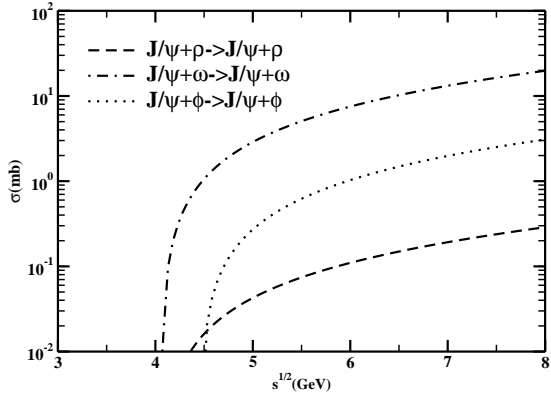}
\caption{Scattering cross section of $J/\psi$ with vector mesons $\rho^0$, $\omega$ and $\phi$
using effective interaction (\ref{Lag}).} 
\label{cross_sn}
\end{center}
\end{figure}

The spin averaged modulus square of 
the total amplitudes corresponding to the amplitudes given above 
are listed in the Appendix. With the help of these 
amplitudes we finally obtain the interaction cross section as a function of
the centre of mass energy $(\sqrt{s})$. 
The  elastic scattering cross sections of $J/\psi$
with vector mesons $\rho^0$, $\omega$ and $\phi$ are plotted against $\sqrt{s}$ 
in Fig.~\ref{cross_sn}.
We find that the cross sections obtained due to the interaction of $J/\psi$ with
$\rho^0$ and $\omega$  have  values within 1 to 10 mb
for $\sqrt{s}\sim 8$ GeV. This is in good 
agreement with the the results provided in Ref.~\cite{Haglin}.

Finally using these amplitudes of the elastic scattering between the charmonia ($J/\psi$ and $\eta_c$) 
with the vector mesons and putting them in Eq.~\ref{transport} we obtain the
drag and diffusion coefficients of the $J/\psi$ and $\eta_c$ in  hadronic matter.

The scattering lengths of charmonia ($J/\psi$ and $\eta_c$) with light hadrons ($\pi$,$\rho^0$ and $N$)
have been studied in the literature in order to estimate the interactions of $J/\psi$ and $\eta_c$
with light hadrons at low energy. 
In ~\cite{Hatsuda} the low-energy interactions of $J/\psi$ as well as $\eta_c$ 
with $\pi$, $\rho^0$ or $N$ have been investigated by Yokokawa et al. in the quenched lattice QCD framework.
From the scattering lengths, $a$ (say) of $J/\psi$ or $\eta_c$ interacting with light
hadrons $H$ (where $H=\pi, \rho^0$ and $N$) we
can extract the dimensionless threshold, the $T$-matrix element
by using the relation
\be
T=4\pi[m_{(J/\psi,\eta_c)}+m_H]a~.
\ee
Using these ${\overline {|T|^2}}$ in place of ${\overline {|M|^2}}$ in Eq.~(\ref{transport}),
we can get an alternative estimation of the diffusion and drag coefficients of $J/\psi$ 
as well as $\eta_c$ mesons in hadronic matter~\cite{SSSJ_B}. 
The extracted values of $T$ from $a$ (in $fm$) are given in Table I.

\begin{table}[h]
\begin{center}
\begin{tabular}{|c|c|c|c|}
\hline
& & & \\
 & $J/\psi\pi$ & $J/\psi\rho^0$ & $J/\psi N$ \\
& & &  \\
\hline
& & & \\
$a ({\rm fm})$ & 0.0119 $\pm$ 0.0039  &  0.23 $\pm$ 0.08  & 0.71 $\pm$ 0.48  \\
& & & \\
T & 2.45 $\pm$ 0.8 & 56.69 $\pm$ 19.71 & 182.60 $\pm$ 123.45  \\
& & &  \\
\hline
 &  &  &   \\
  & $\eta_c\pi$ & $\eta_c\rho^0$ & $\eta_c N$ \\
\hline 
& & & \\
$a ({\rm fm})$ & 0.0113 $\pm$ 0.0035  & 0.21 $\pm$ 0.11  & 0.70 $\pm$ 0.66  \\
& & & \\
T &2.24 $\pm$ 0.69  & 50.20 $\pm$ 26.29 & 174.85 $\pm$ 164.86 \\
& & & \\
\hline
\end{tabular}
\caption{Table showing the extracted values of T-matrix from the
spin averaged values of scattering length, $a$, which are obtained 
in the framework of quenched lattice calculation by 
Yokokawa et. al \cite{Hatsuda}.}
\end{center}
\end{table}
\section{Results}
\begin{figure}
\begin{center}
\includegraphics[scale=0.3]{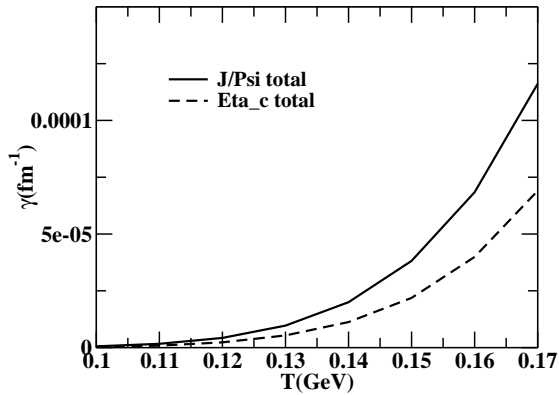}
\caption{The drag coefficient ($\gamma$) as a function of temperature calculated
in the effective Lagrangian approach.} 
\label{drag_EL}
\end{center}
\end{figure}

We begin this section by plotting in Fig.~\ref{drag_EL} the drag coefficients of
the $J/\psi$ (solid line) and $\eta_c$ (dashed line) as a function of temperature.
As mentioned before, the drag is a measure of the momentum transfer between the 
$J/\psi$ (or $\eta_c$) and the thermal hadrons  weighted by the interactions implemented
through $|M|^2$. The average momentum of the bath particles increases with temperature.
Therefore, the thermal hadrons gain the ability to transfer larger momentum through interactions
as the temperature of the bath increases. This causes the rise of  drag  at high temperatures  
both for $\J$ and $\eta_c$. 

\begin{figure}
\begin{center}
\includegraphics[scale=0.3]{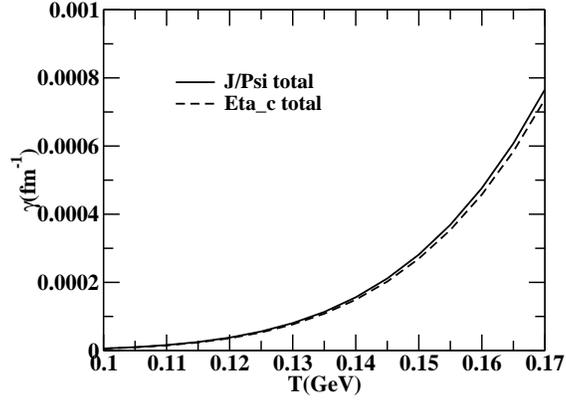}
\caption{The drag coefficient ($\gamma$) as a function of temperature obtained
using scattering lengths.} 
\label{drag_SL}
\end{center}
\end{figure}

In Fig.~\ref{drag_SL} we show the corresponding results for the case where
the amplitudes are extracted from scattering lengths. In this case the  difference 
in drag coefficients between the $\J$ and $\eta_c$ turns out to be insignificant.
This is because the scattering lengths obtained from lattice are of similar magnitude
unlike the effective interaction which is different for the  $\J$ and $\eta_c$.

The drag coefficient estimated by taking the scattering length from Ref.~\cite{Hatsuda}
differs conspicuously from  its value obtained in effective Lagrangian  approach.  This
is anticipated because the calculations performed in the two approaches
involve
very different values of dynamical quantities {\it i.e.}, the scattering cross section.  
A straightforward comparison of the magnitude of the cross sections will help in understanding 
the difference in the value of the drag coefficient. The magnitude of the $J/\psi-\rho$ elastic cross 
section as displayed in Fig.~\ref{cross_sn} is $\sim$ a few tenth of a mb which is
smaller than the value obtained  in ~\cite{Hatsuda} (where the 
pseudoscalars are far from their physical masses and 
coupling strength is presumably larger). 
This indicates that drag estimated within the effective Lagrangian model will be smaller 
than scattering length approach as exhibited in Figs.~\ref{drag_EL} and \ref{drag_SL}.  
However,  it is crucial to note that with either value of the drag coefficient 
the momentum distribution of the $J/\psi$ in the hadronic medium remain 
largely unaltered.

\begin{figure}
\begin{center}
\includegraphics[scale=0.3]{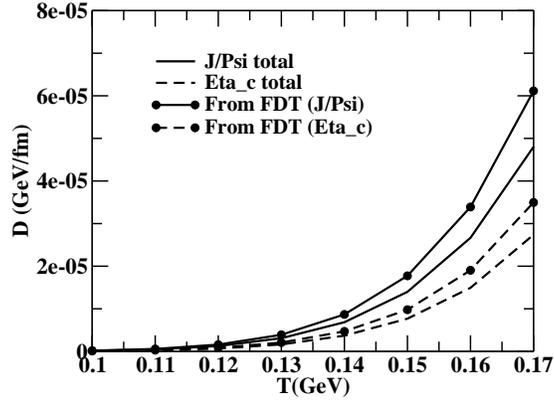}
\caption{The diffusion coefficient ($D$) as a function of temperature calculated
in the effective Lagrangian approach.} 
\label{diff_EL}
\end{center}
\end{figure}

\begin{figure}
\begin{center}
\includegraphics[scale=0.3]{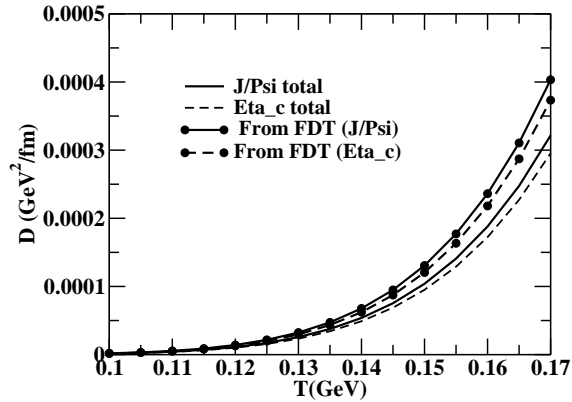}
\caption{The diffusion coefficient ($D$) as a function of temperature obtained
using scattering lengths.} 
\label{diff_SL}
\end{center}
\end{figure}

Now we display our results for the diffusion coefficient, $D$ which is plotted
against $T$ in Figs.~\ref{diff_EL} and \ref{diff_SL}
for the effective Lagrangian and scattering length approaches respectively. 
In addition to the
results of direct calculation using eq.~(\ref{transport}), 
the results from the fluctuation-dissipation theorem (FDT) are
also shown by solid ($J/\psi$) and dashed ($\eta_c$) lines with solid circles  
in Figs.~\ref{diff_EL} and \ref{diff_SL}.
As for the earlier case of
the drag coefficient, the diffusion in the scattering length approach is similar
for the $\J$ and $\eta_c$ mesons. The rise of diffusion coefficients with increasing
temperature has the same origin as that of drag coefficients as explained above. 

The heavy quarks (HQs) momentum suppression have been used to understand the properties of the  QGP  
matter. Several perturbative QCD (pQCD) based analyses (see~\cite{hfr,Rapp:2008tf,Averbeck:2013oga}
and refs. therein)  have been performed to study the 
suppression of HQs at high momentum region, where the pQCD techniques are reliable.  
For an accurate characterization of the QGP it is essential to
discern the role of hadronic matter at the same kinematic domain.
In this spirit the suppression of heavy mesons (open or hidden)
may also be considered at the high momentum domain to estimate and disentangle 
the contributions from the hadronic matter.

The suppression of $\J$ (or $\eta_c$) at high momenta in nuclear collisions
compared to proton-proton collision may be 
approximately estimated as,  $R_{AA}\sim e^{-\Delta\tau\,\gamma}$~\cite{Rapp}, where
$\Delta\tau$ is typically the life time of the hadronic phase. Taking $\Delta\tau\sim 5$  fm/c
and $\gamma\sim 10^{-4}$ one finds that $R_{AA}$ is close to unity. Thus the
hadronic phase does not play a significant role in the suppression of $J/\psi$
(or $\eta_c$)at high $p_T$,
therefore, a significant suppression, if observed experimentally,  
is likely to originate from the
QGP phase of the evolving fireball produced in relativistic heavy ion collisions.
In the earlier investigations~\cite{Juan_D,Rapp,SSSJ_B} 
on $D$ meson diffusion in hadronic matter, our estimations~\cite{SSSJ_B} from 
scattering lengths of $D$ meson with other hadrons are more or less close to 
the estimations of other group~\cite{Juan_D,Rapp} 
[$\gamma({\rm fm}^{-1})\approx$ 0.012-0.032~\cite{Juan_D}, 0.01-0.03~\cite{Rapp},
0.005-0.027~\cite{SSSJ_B} within hadronic temperature domain $T=120-170$ MeV].
On this basis, our estimations for $J/\psi$ are expected to be similar with
other models.

\section{Summary and discussions}
We have estimated the drag and diffusion coefficients of $J/\psi$ and $\eta_c$ in a hot
hadronic medium using effective field theory and $T$ matrices obtained within the ambit of 
quenched lattice QCD calculations.  The values of these transport coefficients turn out 
to be small compared to the values obtained for open charmed hadrons for the temperature range
relevant for the hadronic phase expected to be formed in the later stages of the evolving matter
produced in nuclear collisions at RHIC and LHC energies.  We find that the suppression 
of $\J$ and $\eta_c$ at high momentum may not be significant in the hadronic phase. This finding
prevails with the values of the drag and diffusion coefficients obtained with inputs
either from lattice QCD or effective Lagrangian approach {\it i.e.} the main conclusion
of this work is not affected by the variation in the values of drag obtained within the ambit
of the two approaches adopted here. 

Therefore, such a suppression
if observed experimentally will possibly indicate the creation of QGP in heavy ion collisions at 
relativistic energies. 

Of late the shear viscosity ($\eta$) to entropy density ($s$)  ratio ($\eta/s$) 
has been considered as an useful quantity
to characterize the matter formed in heavy ion collisions at  RHIC and LHC 
energies. The value of this ratio is found to be small from the analysis 
of experimental data~\cite{etabys}, which led to the conclusion that the 
matter formed in these collision behaves like a perfect liquid. 

In the present work we find that the momentum diffusion of $J/\psi$ is small 
{\it i.e.} the transfer of momentum between $J/\psi$ and hadrons is inefficient
which means that the shear viscosity is large~\cite{sm}. This again indicates that the 
role of hadrons in the suppression of $J/\psi$ in the high momentum domain
is inconsequential compared to QGP. 

It may be mentioned that the role of the nucleons has been neglected 
in evaluating the drag and diffusion coefficients 
of $J/\psi$ and $\eta_c$ within the ambit of effective Lagrangian approach here.
However,  the inclusion of the nucleons will not change the conclusion
because we found that an enhancement of the drag coefficient, $\gamma$ by a factor of 4 will suppress
the $p_T$ spectra by  a factor less than $1\%$. 

\section{Appendix}

The modulus square of the spin averaged total amplitude
for the processes of $\J + V\rightarrow \eta_c\rightarrow\J +V$
is given by the following expression,
\bea
\overline{|M|^2}=\overline{|M_s|^{2}}+\overline{|M_u|^{2}}+2\overline{M_s M_u^{*}}
\label{M2_J}
\eea
where the respective terms in the expression are given by
\begin{eqnarray}
\overline{|M_s|^2}&=&\frac{g^4_{JV\eta_c}}{36(s-m^2_{\eta_c})^2}\lambda^2({s,m^2_{J/\psi},m^2_{V}})
\nonumber\\
\overline{|M_u|^2}&=&\frac{g^4_{JV\eta_c}}{36(u-m^2_{\eta_c})^2}\lambda^2({u,m^2_{J/\psi},m^2_{V}})
\nonumber\\
\overline{M_sM_u^{*}}&=&\frac{g^4_{JV\eta_c}}{9(s-m^2_{\eta_c})(u-m^2_{\eta_c})}I~.
\nonumber
\end{eqnarray}

where,
\begin{eqnarray}
I&=&\frac{1}{8}[m_J^8+s^4+2s^3(t-2m_V^2)+2sm_V^4(t-2m_V^2)
-4m_J^6(s+m_V^2)+m_V^4(t^2
\nonumber\\
&&+m_V^4)+s^2(t^2 -4 tm_V^2+6m_V^4)
+m_J^4\{6s^2+t^2+6m_V^4+2s(t+2m_V^2)\}
\nonumber\\
&&-2m_J^2\{2s^3+2s(t-m_V^2)(s+m_V^2)
+m_V^2(t^2+2m_V^4)\}]\nonumber.
\end{eqnarray}
and  $\lambda(x,y,z)=x^2+y^2+z^2-2xy-2yz-2zx$ is the triangular function.
Next, the spin averaged modulus square of total amplitude
for the processes $\eta_c+V \rightarrow J/\psi \rightarrow\eta_c +V$
are given by
\begin{equation}
\overline{|M|^2}=\overline{|M_s|^{2}}+\overline{|M_u|^{2}}+2\overline{M_s M_u^*}
\end{equation}
where
\bea
\overline{|M_s|^2}&=&(g^{4}_{JV\eta_c}/3)\left\{\frac{s}{4}(t-4m_V^2)(s+m_V^2-m^2_{\eta_c})^2
+\frac{1}{8}(s+m_V^2-m^2_{\eta_c})^4
\right.\nn\\
&&\left.+\frac{s^2}{4}(t^2-4tm^2_V+8m^4_V)\right\}/(s-m_J^2)^2~.
\eea
\bea
\overline{|M_u|^2}&=&(g^{4}_{JV\eta_c}/3)\left\{\frac{u}{4}(t-4m_V^2)(u+m_V^2-m^2_{\eta_c})^2
+\frac{1}{8}(u+m_V^2-m^2_{\eta_c})^4
\right.\nn\\
&&\left.+\frac{u^2}{4}(t^2-4tm^2_V+8m^4_V)\right\}/(u-m_J^2)^2.
\eea
and
\begin{eqnarray}
\overline{M_s M_u^*}&=&(g^{4}_{JV\eta_c}/3) \left[m_{\eta_c}^8 + m_V^8 + s^4 
- 4 m_{\eta_c}^6 (m_V^2 + s)+2 m_V^4 s (3 s - t)
\right.\nn\\
&&\left.  + 2 s^3 t - s t^3 + m_V^6 (-4 s + 2 t)-2 m_V^2 s (2 s^2 + s t - 2 t^2)
\right.\nn\\
&&\left. 
+ 2 m_{\eta_c}^4 \{3 m_V^4 + m_V^2 (2 s + t)+s(3 s + t)\}
-2 m_{\eta_c}^2\{2 m_V^6 - 2 m_V^2 s (s - 2 t) 
\right.\nn\\
&&\left. - 2 m_V^4 (s - t)
+s (2 s^2 + 2 s t - t^2)\}\right]/8(s-m_J^2)(u-m_J^2).
\end{eqnarray}

{\bf Acknowledgment:} 
S.G. is supported by FAPESP, Grant No 2012/16766-0.
S.K.D  acknowledges the support by the ERC StG under the QGPDyn
Grant No 259684.

\end{document}